\documentclass[namedreferences,hyperref,optionalrh,solaromanenum]{spr-sola}

\usepackage{graphicx}                    
\usepackage{amsmath}  
\usepackage{bm}       
\usepackage{arydshln}
\usepackage{color}                       
\newcommand{\EQ}{\begin{equation}}
\newcommand{\EN}{\end{equation}}
\newcommand{\EQA}{\begin{eqnarray}}
\newcommand{\ENA}{\end{eqnarray}}
\newcommand{\nab}{\mbox{\boldmath $\nabla$} {}}
\newcommand{\brac}[1]{\langle #1 \rangle}

\newcommand{\DIV}{\bm{\nabla} \cdot }

\newcommand{\AAA}{\bm{{A}}}
\newcommand{\BB}{\bm{{B}}}
\newcommand{\JJ}{\bm{{J}}}
\newcommand{\UU}{\bm{{U}}}

\newcommand{\uu}{\bm{{{u}}}}

\newcommand{\ff}{\bm{{{f}}}}
\newcommand{\kk}{\bm{{{k}}}}
\newcommand{\xx}{\bm{{{x}}}}

\newcommand{\eee}{\hat{\mbox{\boldmath $e$}} {}}

\def\kf{k_\mathrm{f}}
\def\ii{{\rm i}}

\def\const{{\rm const}}

\def\Pm{P_\mathrm{m}}
\def\Rm{R_\mathrm{m}}

\newcommand{\mean}[1]{\overline{#1}}

\newcommand{\cs}{c_{\rm s}}
\newcommand{\urms}{u_{\rm rms}}

\def\onethird{{\textstyle{1\over3}}}
\def\onehalf{{\textstyle{1\over2}}}

\def\onethird{{\textstyle{1\over3}}}

\newcommand{\RRRR}{\mbox{\boldmath ${\sf R}$} {}}
\newcommand{\SSSS}{\mbox{\boldmath ${\sf S}$} {}}

\newcommand{\Fig}[1]{Figure~\ref{#1}}

\newcommand{\BoldVec}[1]{\mathchoice%
  {\mbox{\boldmath $\displaystyle     #1$}}%
  {\mbox{\boldmath $\textstyle        #1$}}%
  {\mbox{\boldmath $\scriptstyle      #1$}}%
  {\mbox{\boldmath $\scriptscriptstyle#1$}}%
}

\usepackage{subfig}

\chardef\us=`\_

\begin{document}

\begin{frontmatter}

\title{Hysteresis near the transition of the large-scale dynamo in the presence of the small-scale dynamo}

%
\author[addressref={aff1},email={vindyavashishth.rs.phy19@itbhu.ac.in}]{\inits{V.}\fnm{Vindya}~\snm{Vashishth}}


\runningauthor{V. Vashishth}
\runningtitle{\textit{Large-scale dynamo in the presence of the small-scale dynamo}}

\address[id=aff1]{Department of Physics, Indian Institute of Technology (BHU) Varanasi,  Uttar Pradesh, India}

\begin{abstract}
In Sun and solar-type stars, 
there is a critical dynamo number for the operation of a large-scale dynamo, below which the dynamo ceases to operate. This region is known as the subcritical region. 
Previous studies showed the possibility of operating the solar-like large-scale (global) dynamo in the subcritical region without a small-scale dynamo. As in the solar convection zone, 
both large- 
and small-scale 
dynamos are expected to operate at the same time and location, we check the robustness of the previously identified subcritical dynamo branch in a numerical model in which both large- and small-scale dynamos are excited. 
For this, we use the {\sc Pencil Code} and set up an $\alpha\Omega$ dynamo model with uniform shear and helically forced turbulence. We have performed a few sets of simulations at different relative helicity to explore the generation of large-scale oscillatory fields in the presence of small-scale dynamo. We find that in some parameter regimes, the dynamo shows hysteresis behavior, i.e., two dynamo solutions are possible depending on the initial parameters used. A decaying solution when the dynamo was started with a weak field and a strong oscillatory solution if the dynamo was initialized with a strong field.  
Thus, the existence of the sub-critical branch of the large-scale dynamo in the presence of small-scale dynamo is established. 
However, the regime of hysteresis is quite narrow with respect to the case without the small-scale dynamo. 
Our work supports the possible existence of large-scale dynamo in the sub-critical regime of slowly rotating stars. 
\end{abstract}

%

\end{frontmatter}


\section{Introduction}\label{s:1}
The Sun exhibits magnetic fields and cycles of remarkable complexity. Unlike clockwork, its magnetic behavior is not strictly periodic; instead, it demonstrates a dynamic interplay of various characteristics, including cycle duration and amplitude, which vary from one cycle to the next \citep{Hat15, KMB18}. At times, the Sun enters phases of grand minima, characterized by extended periods of low magnetic activity. 
Analyses of radiocarbon data of last several thousand years
unveiled several such instances  \citep{ USK07, Uso13, Biswas23}.

This intricate dance of magnetic fields and cycles finds its origins in the mechanism of the large-scale dynamo \citep{Mof78}. 
This process lays the foundation for the creation and sustenance of the magnetic fields that envelop the Sun.
The large-scale dynamo is the foundational mechanism  
behind generating the magnetic field.
It operates globally in the solar convection zone and establishes substantial magnetic fields extending over the surface.
This dynamo action is powered by helical convection and differential rotation in the solar convection zone. 
This is because the toroidal field is generated through the stretching of the poloidal field by the differential rotation, known as the  $\Omega$ effect. This toroidal field gets converted into the poloidal one via helical flow formally known as $\alpha$ effect \citep{Pa55, Steenbeck1966}.
There is another additional mechanism for the generation of the poloidal field in the Sun known as the Babcock--Leighton process \citep[see][for recent reviews on dynamo modelling using this process]{Cha20, Kar23}.


Observations have revealed an intriguing type of magnetic field that persists even in the quiet phase of the sun, which is commonly referred to as a small-scale, fluctuating, turbulent, or internetwork field—as it exists in the internetwork regions. 
This field exists in mixed polarity even at the
resolution limits of present-day instruments. Despite the longstanding awareness of the existence of this magnetic field  (e.g., \citet{FS72,S12}), a detailed understanding of its nature and origin has remained an ongoing challenge \citep{Rempel2023}.
The presence of this magnetic field rises from the mechanism
referred to as the small-scale (local) dynamo. 
This process involves amplifying a seed magnetic field through repeated random stretching, bending, and folding within a sufficiently random three-dimensional velocity field, all without net helicity.

There is a possibility that the quiet-Sun magnetic field could be attributed to a large-scale global dynamo. This concept suggests that the disintegration of a large-scale magnetic field could generate a small-scale magnetic field by transferring magnetic energy to smaller scales. Additionally, it can be argued that the decay of active regions might contribute to forming small-scale magnetic fields \citep{STB87,deW05, S12, KB16}. Nonetheless, it is worth noting that none of these arguments can be conclusively affirmed. The small-scale magnetic field exhibits characteristics that do not align with the solar cycle, as it lacks a significant correlation with the larger-scale global magnetic cycle. Moreover, it displays no latitudinal variation, as demonstrated by various studies (e.g., \citet{HST03,Sa03,L08,Li11,BLS13,JW15b}).
The purpose of the present study is to explore the bistability of the large-scale dynamo in the presence of a small-scale dynamo with an $\alpha\Omega$ dynamo.

In the $\alpha\Omega$ dynamo model, there exists a critical parameter known as the critical dynamo number below which the magnetic field ceases to operate, resulting in a decaying field \citep{Pa55, Choudhuri_Book,BS05}.
This regime is referred to as the subcritical dynamo phase. Conversely, when the dynamo number exceeds this critical value, the system enters the supercritical dynamo phase, characterized by sustained magnetic field generation. This dynamo transition has been extensively documented in studies focusing on large-scale dynamo \citep{Choudhuri_Book}. 
It is interesting to know in what regime the solar dynamo operates. Observations \citep{R84, Met16} and dynamo modeling \citep{KN17, Vindya21, KKV21, CS17, Vindya23, Ghosh2024} hint that the solar dynamo is possibly operating near the critical dynamo transition or at least not in highly supercritical reguime.
Recent investigations using
mean-field modelling \citep{KO10,Vindya21} and turbulent numerical simulations \citep{KKB15, Oliveira21} have unveiled intriguing phenomena suggesting that the dynamo process can persist even in subcritical regions. This behavior is manifested through hysteresis, a phenomenon observed in the context of large-scale dynamo dynamics. 
Complementing these findings, \citet{PPF2022}, underscores the effectiveness of nonlinear optimization—previously utilized for identifying minimal disturbances in shear flows—as a potent numerical approach for methodically probing subcritical dynamo actions in electrically conducting flows.

The aforementioned studies did not capture the operation of the small-scale dynamo, which is ubiquitous in solar/stellar convection zones. It is obvious that the operation of the large-scale dynamo is affected by the small-scale dynamo-generated field. 
While some studies have explored the interaction of small-scale dynamo on large-scale one \citep[e.g.,][]{KB16,Bhat16}, here we are interested in the possibility of 
dynamo hysteresis behavior of the large-scale magnetic field in the presence of small-scale dynamo 
to demonstrate the robustness of the operation of the subcritical dynamo.


\section{Model}\label{s:2}

Following the works of \citet{Kar15, KB16}, we build our theoretical $\alpha\Omega$ dynamo model assuming an isothermal and compressible environment. The pressure in this medium is characterized by the equation of state $p=\cs^2 \rho$, where $\cs$ represents the constant speed of sound, and $\rho$ signifies the density. The fundamental equations governing this model are:
\begin{equation}
\frac{D \UU}{D t} = -S U_x \hat y -c_{\rm s}^2\nab\ln\rho + \rho^{-1} \left[\JJ \times \BB + \nab\!\cdot(\!2\rho\nu\SSSS)\right] + \ff,
\end{equation}
\begin{equation}
\frac{D\ln\rho}{D t} =-\nab\cdot\UU,
\end{equation}
\begin{equation}
\frac{\partial \AAA}{\partial t} + \mean \UU^{(S)} \cdot \nab \AAA = - S A_y \hat x + \UU\times\BB +  \eta \nab^2 \AAA.
\end{equation}

In these equations, $D/Dt$ denotes the advective time derivative, expressed as $D/Dt = \partial/\partial t + (\UU  + \mean \UU^{(S)})\cdot \bm\nab$. The term $\mean \UU^{(S)}=(0,Sx,0)$ with $S=\const$ represents the large-scale externally applied uniform shear flow. Other parameters include the magnetic vector potential $\AAA$, the magnetic field ${\BB} = \nab\times {\AAA}$, the microscopic diffusivity $\eta$, the kinematic viscosity $\nu$, the current density $\JJ = \mu_0^{-1} \nab\times\BB$, and a specific forcing function $\ff$.

The traceless rate of the strain tensor $\SSSS$ is represented by
${\sf S}_{ij} = \onehalf (U_{i,j}+U_{j,i}) - \onethird \delta_{ij} \DIV \bm{U},$ 
ignoring the minimal contribution of $\mean \UU^{(S)}$.
In this equation, the commas denote partial differentiation with respect to the coordinate ($i$ or $j$).

Turbulence in this environment is sustained by supplying energy 
to the system through a
helical and temporally random in time ($\delta$-correlated) forcing function
$\BoldVec{f} = \BoldVec{f}(\xx,t)$. 
This forcing function is defined as
\EQ
\ff(\xx,t)={\rm Re}\{N\ff_{\kk(t)}\exp[\ii\kk(t)\cdot\xx+\ii\phi(t)]\}.
\label{ForcingFunction}
\EN
Here, $\xx$ denotes the position vector, $\kk(t)$ is a random wavevector chosen at each timestep from a certain range of many possible wavevectors, and the phase $-\pi<\phi(t)\le\pi$ also varies randomly at each timestep.
On dimensional grounds, we choose $N=f_0 c_{\rm s}(|\kk|c_{\rm s}/\delta t)^{1/2}$,
where $f_0$ is a dimensionless forcing amplitude.

The generation of transverse helical waves is facilitated through the utilization of Fourier amplitudes
\citep{Hau04},
\begin{equation}
\ff_{\kk}=\RRRR\cdot\ff_{\kk}^{\rm(nohel)}\quad\mbox{with}\quad
{\sf R}_{ij}={\delta_{ij}-\ii\sigma\epsilon_{ijk}\hat{k}_k
\over\sqrt{1+\sigma^2}},
\label{eq: forcing}
\end{equation}
where $\sigma$ represents the degree of helicity in the forcing, with $\sigma=1$ indicating the highest positive helicity. 
The formulation of the non-helical forcing function is represented with,
$\ff_{\kk}^{\rm(nohel)}=
\left(\kk\times\eee\right)/\sqrt{\kk^2-(\kk\cdot\eee)^2}, \nonumber$
where $\eee$ is an arbitrary unit vector that is not in alignment with 
$\kk$. Note that $|\ff_{\kk}|^2=1$ and
$\ff_{\kk}\cdot(\ii\kk\times\ff_{\kk})^*=2\sigma k/(1+\sigma^2)$.

The fluid and magnetic Reynolds numbers and the magnetic Prandtl number are defined as
\begin{equation}
 R_e = \frac{\urms}{\nu\kf}, \quad
\Rm = \frac{\urms}{\eta\kf},\quad
\Pm=\frac{\nu}{\eta}, \quad
\end{equation}
where 
$\urms =\langle \uu^2 \rangle^{1/2}$ is the root-mean-square (rms) value of the velocity in the statistically stationary state.
Here, $\langle\cdot\rangle$ indicates the averaging across the entire domain, and $\kf$ is the average forcing wavenumber. 
To investigate the small-scale dynamo effects alongside the large-scale dynamo, we maintain a large value of 
$R_m$ and dynamo number, $D$ which is defined as  
\begin{equation}
D = C_\alpha C_\Omega \quad {\rm where}, 
~C_\alpha = \frac{\alpha_0}{\eta_{T0} k_1} \quad {\rm and} 
~~C_\Omega = \frac{|S|} {\eta_{T0} k_1^2},\quad
\end{equation} 
as elaborated in Table 1. 
Here, $\alpha_0 = -\tau \brac{\omega \cdot \bm u}/3$, $\eta_{T0}$ is the total magnetic diffusivity and is given by $\eta_{T0} = \eta + \eta_{t0}$, with $\eta_{t0} = \tau (\brac{\bm u}^2)/3 $, and $\tau = (u_{\rm rms} k_f)^{-1}$.
In this work, we have followed Run IV of \cite{KB16}, which excites both the large-scale and small-scale dynamos.

To establish a connection with solar/stellar convection zones, we imagine a 3D box positioned at the northern hemisphere of the sun, with dimensions of $L_x = L_y = L_z = 2 \pi$. In this configuration, $x, y, z$ coordinates correspond to the radially outward, azimuthal (toroidal), and latitudinal directions, respectively. 
The boundary conditions implemented in our model are shearing--periodic along the $x$-axis and simple periodic along the 
$y$ and $z$ axes.
For all of our simulations, we consistently use
$S=-0.05$, $f_0=0.01$, and $\kf = 3 k_1$, where $k_1=2\pi/L_x = 1$ represents the smallest wavenumber achievable in the given spatial domain.
In terms of units, we adopt a non-dimensional approach by assigning the values $\cs=\rho_0=\mu_0=1$, where
$\rho_0 = \brac{\rho}$ is the time-invariant volume-averaged density and $\mu_0$ is the magnetic permeability.
For the initial conditions, we set both $\uu$ and $\ln\rho$ to zero and introduce a small-scale Gaussian noise with a low amplitude of $10^{-4}$ into the magnetic vector potential.
All numerical simulations in this study were conducted using the {\sc Pencil Code}
\footnote{\url{http://github.com/pencil-code}} \citep{pencil_code}.
The grid resolution of all runs presented in this paper is $144\times144\times144$.

\section{Results $\&$ Discussion}\label{s:3}

\begin{figure}
\centering
\includegraphics[scale=0.34]{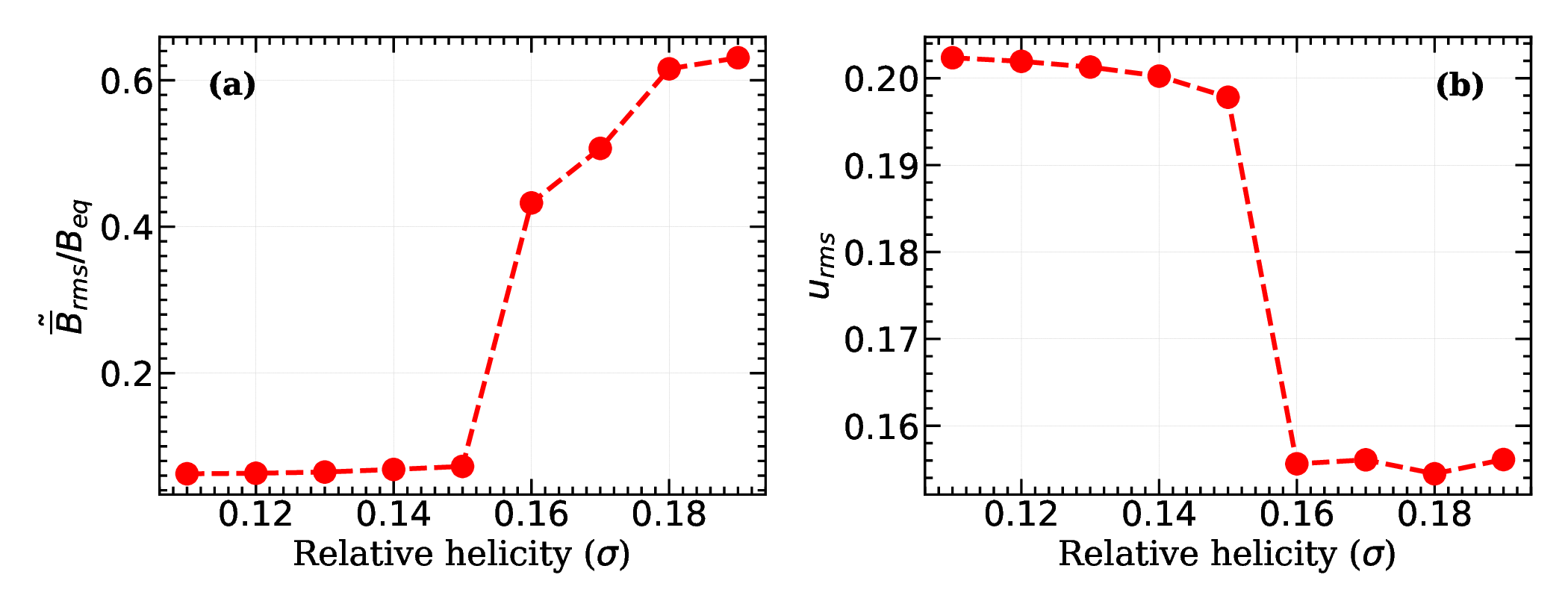} 
 \caption{(a) Normalized temporal mean of rms value 
 of the large-scale magnetic field and (b) temporal mean of $u_{\rm rms}$ as a function of $\sigma$.
 }
   \label{fig:dynamotransion}
\end{figure}

In our study, we conducted a series of simulations by systematically altering the helicity parameter $\sigma$, which 
is the characteristic parameter of the turbulent forcing.  
We examined the large-scale magnetic field by evaluating the quantity, $\overline{B}_{\rm rms}$, which is the temporal mean in the stationary state of the large-scale field over the whole domain and is formulated as, $\Tilde{\overline{B}}_{\rm rms} =\langle\langle B_x\rangle_y{^2} + \langle B_y\rangle_y{^2} + \langle B_z\rangle_y{^2}\rangle^{1/2}_{xzt}$. The small-scale field is defined as the residual of the total and the large-scale quantities and thus defined as
\textbf{$\overline{\bm b^2} =\overline{ \bm B^2}-\overline{\bm B}^2$. }
Our findings, as illustrated in \Fig{fig:dynamotransion}, demonstrate the temporal mean of the large-scale magnetic field, denoted as $\Tilde{\overline{B}}_{\rm rms}$ which we normalized with $B_{eq} (= u_{\rm rms})$, which is the volume-averaged equipartition magnetic field. This normalization provides a context for the background magnetic field. 
The set of parameters used in these simulations is summarized in Table 1. Our analysis indicates that for values of the helicity parameter $\sigma$ lower than approximately 0.16, dynamo activity is notably absent, as indicated by the absence of a large-scale magnetic field (see Table 1).

\begin{table}
\caption{Summary of all of the runs starting with weak seed field are listed along with the control parameters $R_e, R_m, u_{\rm rms}, \sigma$. The table also contains the run time ($T_{r}$ (in diffusion time scale),  normalized temporal mean of the large-scale field  $\Tilde{\overline{B}}_{\rm rms}$ and small-scale field, $\Tilde{\overline{b}}$, and their ratio. For Runs A--I, $k_f=3.1$ and $\nu = 5 \times 10^{-3}$, while in Run F\textquotesingle, $k_f=5.1$ and $\nu = 10^{-2}$. For all the runs, $P_m = 5$. S/D denotes a stable or decaying solution.
}
\label{Table}
\begin{tabular}{ccccccccccc}     
\hline                     
Run& $\sigma$ & $T_{r}$ & $u_{\rm rms}$  & $R_e$ & $R_m$ & $D$ & $\Tilde{\overline{B}}_{\rm rms}$ & $\Tilde{\overline{b}}$ & $\frac{\Tilde{\overline{B}}_{\rm rms}} { \Tilde{\overline{b}}} $ &  S/D \\
\hline
A & 0.10 & 10 & 0.202 & 13.05 & 65.27 & 116.17 & 0.062 & 0.032 & 1.93 & D\\
B & 0.12 & 10 & 0.201 & 13.02 & 65.14 & 127.25 & 0.063 & 0.034 & 1.85 & D\\
C & 0.13 & 10 & 0.201 & 12.98 & 64.92 & 138.76 & 0.064 & 0.044 & 1.45 & D\\
D & 0.14 & 10 & 0.200 & 12.91 & 64.59 & 150.99 & 0.068 & 0.057 & 1.19  & D\\
E & 0.15 & 20 & 0.197 & 12.76 & 63.81 & 165.77 & 0.072 & 0.059 & 1.22 & D\\
F & 0.16 & 200 & 0.155 & 10.04 & 50.20 & 285.71 & 0.432 & 0.280 & 1.54 & S\\
G & 0.17 & 10 & 0.156 & 10.06 & 50.34 & 301.79 & 0.507 & 0.284 & 1.78 & S \\
H & 0.18 & 10 & 0.154 & 9.96  & 49.83 & 326.21 & 0.615 & 0.294 & 2.09 & S\\
I & 0.19 & 10 & 0.156 & 10.07 & 50.35 & 337.15 & 0.630 & 0.294 & 2.14 & S\\
\hdashline
F\textquotesingle & 0.16 & 20 & 0.1015 & 1.983 & 9.915 & 120.61 & 2.773  & 0.376 & 4.65 & S\\
\hline
\end{tabular}
\end{table}

\begin{figure}%
    \centering

    \subfloat[\centering]{{\includegraphics[scale=0.138]{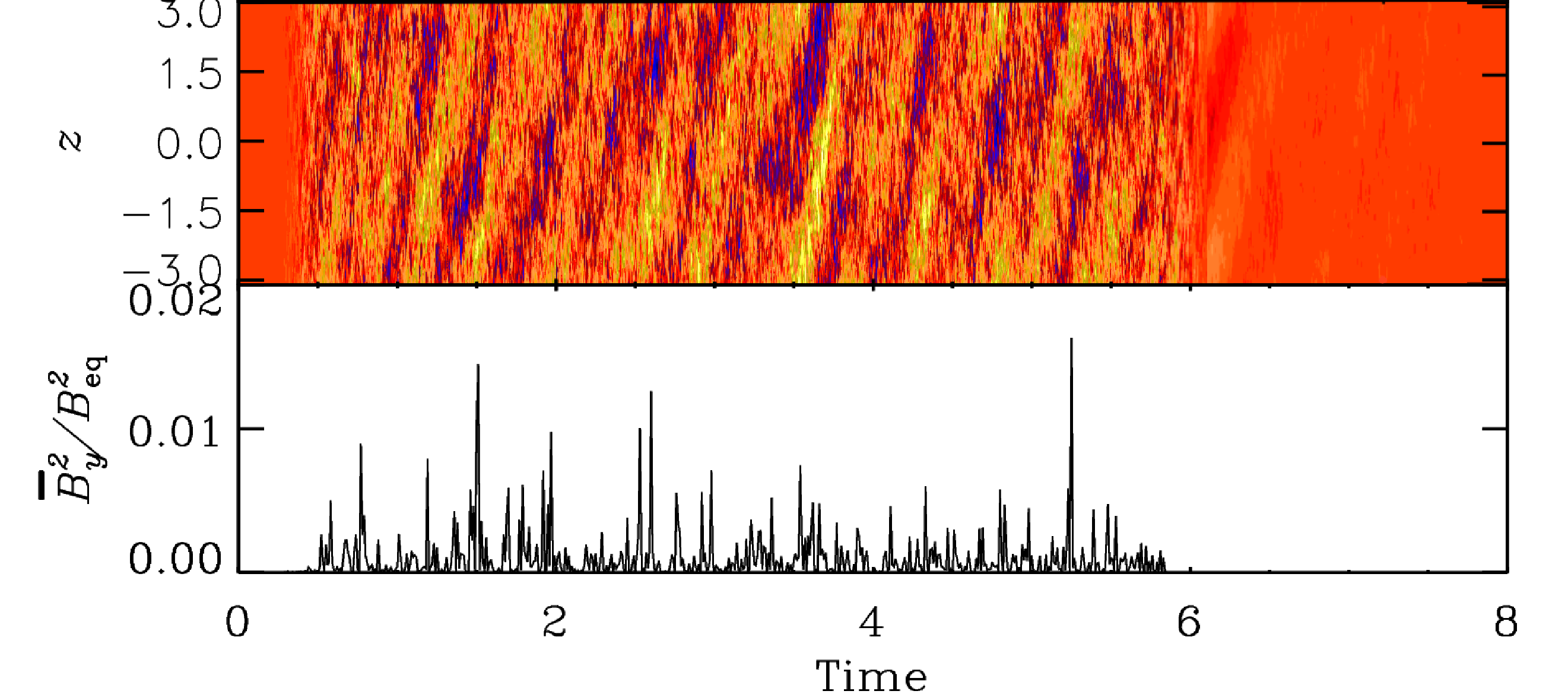} }}
    \hspace*{-0.8 cm}
    \subfloat[\centering]{{\includegraphics[scale=0.270]{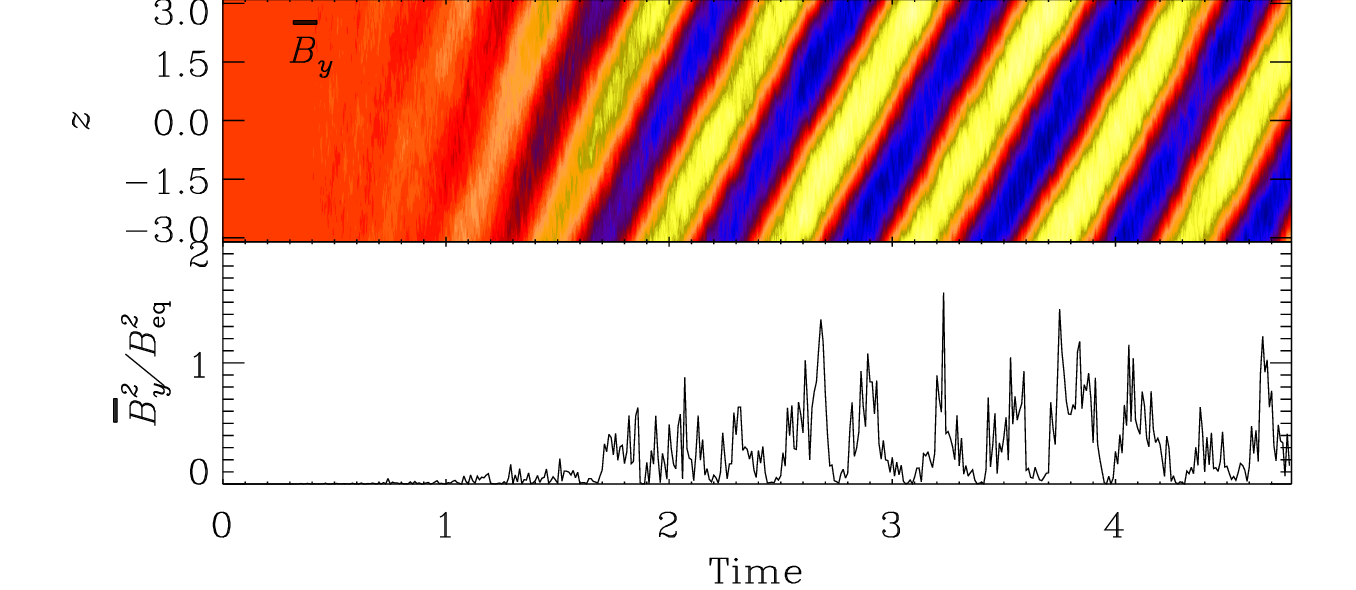} }}
    \caption{Top: Butterfly diagrams of the large-scale magnetic field, $\overline{B_y} (\pi, z, t)$.
    Bottom: 
    Time series plots of $\overline{B_y} (x, z, t)$ taken from an arbitrarily chosen mesh point as a function of time normalized by the diffusive time $(k_{1}^2\eta)^{-1}$.
These results are
from a simulation initiated with a weak seed field at (a) $\sigma = 0.15$ (Run E, Subcritical) and (b) $\sigma = 0.16$ (Run F, Critical).}%
    \label{fig:lsd}%
\end{figure}

\begin{figure}%
    \centering
    \subfloat[\centering]{{\includegraphics[scale=0.276]{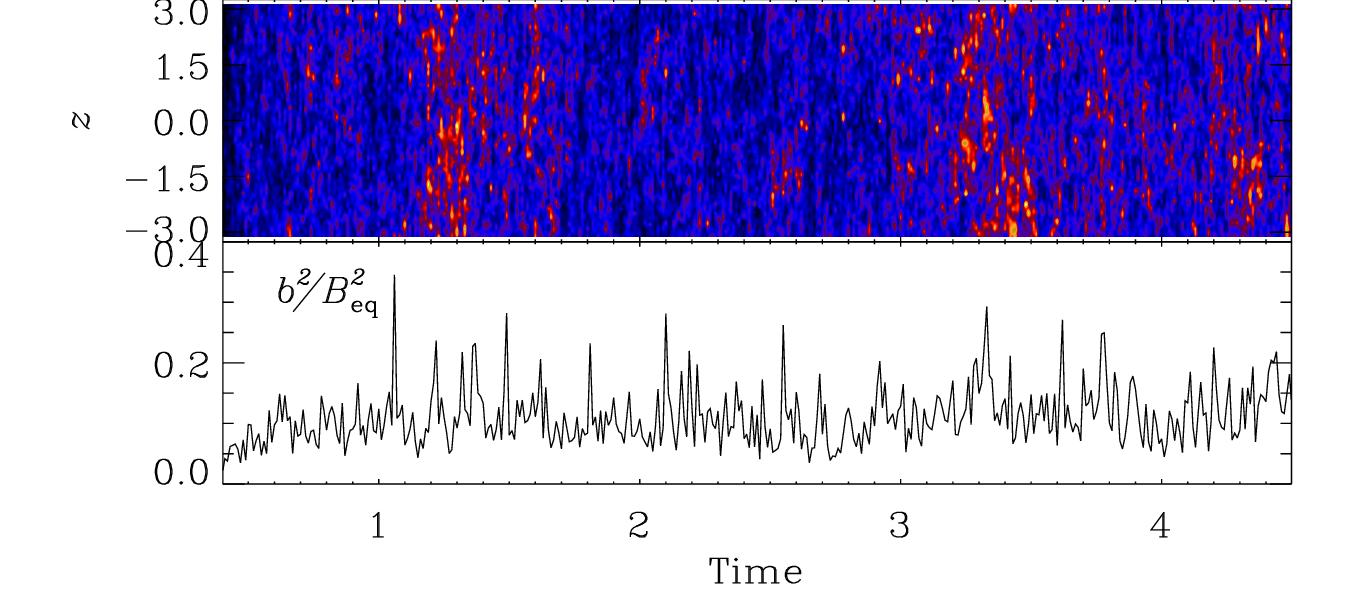} }}
    \hspace*{-0.8 cm}
     \subfloat[\centering]{{\includegraphics[scale=0.276]{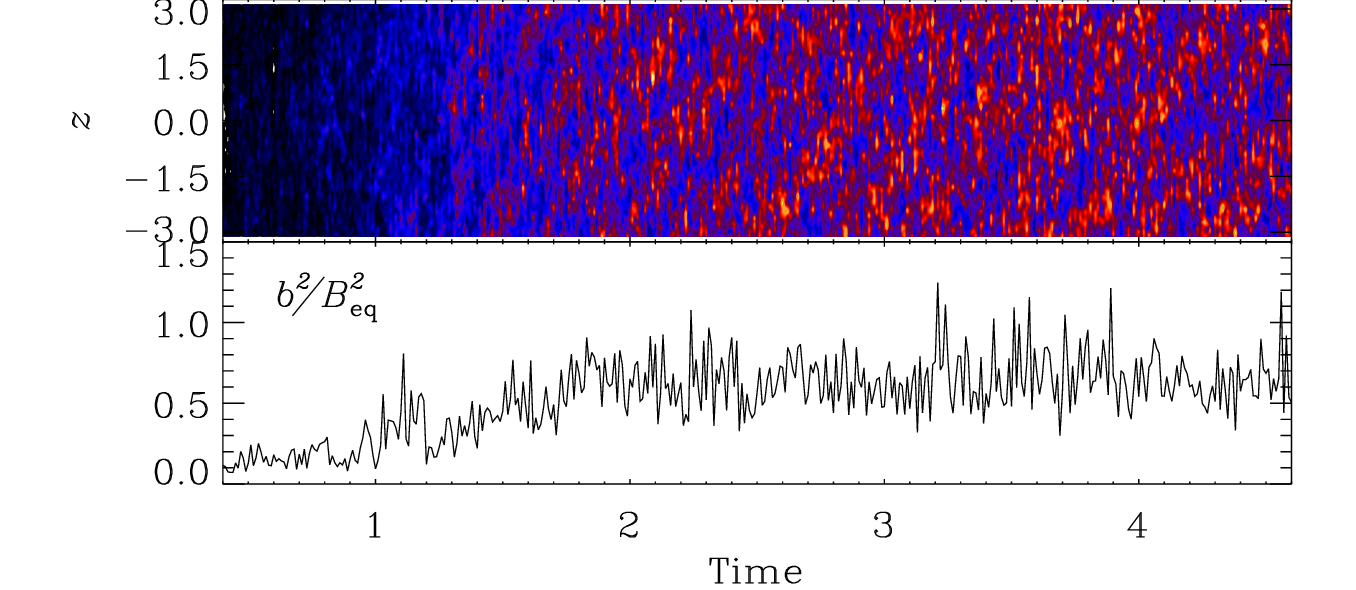} }}
    \caption{Top: butterfly diagrams of the small-scale magnetic field, $\bm b^2 (\pi,z,t)$. Bottom: time series plots of $\bm b^2 (x,z,t)$ taken from an arbitrarily chosen mesh point as
a function of time normalized by the diffusive time scales. These results are
from a simulation initiated with a weak seed field at (a) $\sigma = 0.15$ (Run E, Subcritical) and (b) $\sigma = 0.16$ (Run F, Critical).}%
    \label{fig:ssd}%
\end{figure}

For Run E, $\sigma = 0.15$, as illustrated in \Fig{fig:lsd}a, we demonstrate the spatial-temporal behavior of the mean magnetic field's $y$-component
, $B_y$ (corresponding to the toroidal field in spherical coordinates) and the corresponding time series of $B_y^2$ at a chosen mesh point, normalized by $B_{eq}^2$ (corresponding to the measurements compared with the solar magnetic cycle). 
Notably, there are no distinct magnetic oscillations observed in this scenario. Although a few cycles emerge in the initial time stage, they are short-lived, and the overall magnetic field remains weak. This weak large-scale field disappears after some time, as confirmed by running it for a longer time. A slight increase in $\sigma$ leads to a dynamo transition at $\sigma \approx 0.16$, with the magnetic field becoming significantly stronger than the background ($\Tilde{\overline{B}}_{\rm rms} > B_{eq}$), suggesting a critical $\sigma$ value near 0.16 for dynamo activity.
\Fig{fig:lsd}b depicts the spatial-temporal variation in this case, revealing clear magnetic cycles and dynamo wave propagation in the positive $z$-direction.
\Fig{fig:ssd}(a) and (b), present the spatial-temporal behavior of the small-scale field ($\bm b^2$) normalized with $B_{eq}^2$ at $\sigma = 0.15$ (Run E) and $\sigma = 0.16 $ (Run F), respectively.

Disentangling the origin of these variations is challenging, as the small-scale magnetic field in our simulation arose from the activity of both the small-scale dynamo, represented as $b^2_{SSD}$ and the entanglement with the large-scale field, denoted as $b^2_{tang}$.
These quantities are closely interconnected and evolve in tandem with the development of the large-scale field.
In the early stages, when the large-scale field is still emerging, the small-scale field is predominantly driven by the small-scale dynamo, resulting in $\bm b^2$ being approximately equivalent to $b^2_{SSD}$. Subsequently, as the large-scale field becomes more pronounced, an observable increase in $\bm b^2$ is largely attributed to the tangling effects of the large-scale field. However, upon reaching a specific threshold of $B^2$, $b^2_{tang}$ tends to saturate.

\begin{figure}

\centering
 
\includegraphics[scale=0.5]{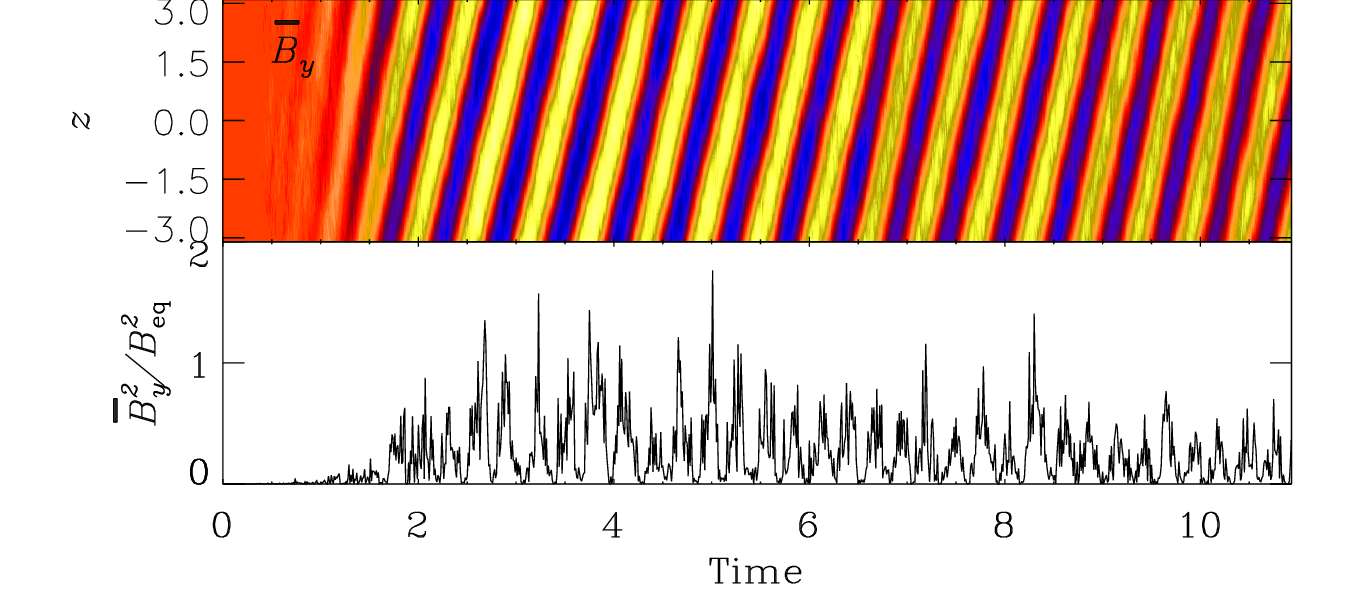} 
 \caption{Butterfly diagram the large-scale magnetic field, $\overline{B_y} (\pi, z , t)$ when the simulation started with a strong magnetic field for the subcritical dynamo Run E ($\sigma = 0.15$).}
   \label{fig:strong_lsd}
\end{figure}

\begin{figure}%
    \centering
    \subfloat[\centering]{{\includegraphics[scale=0.274]{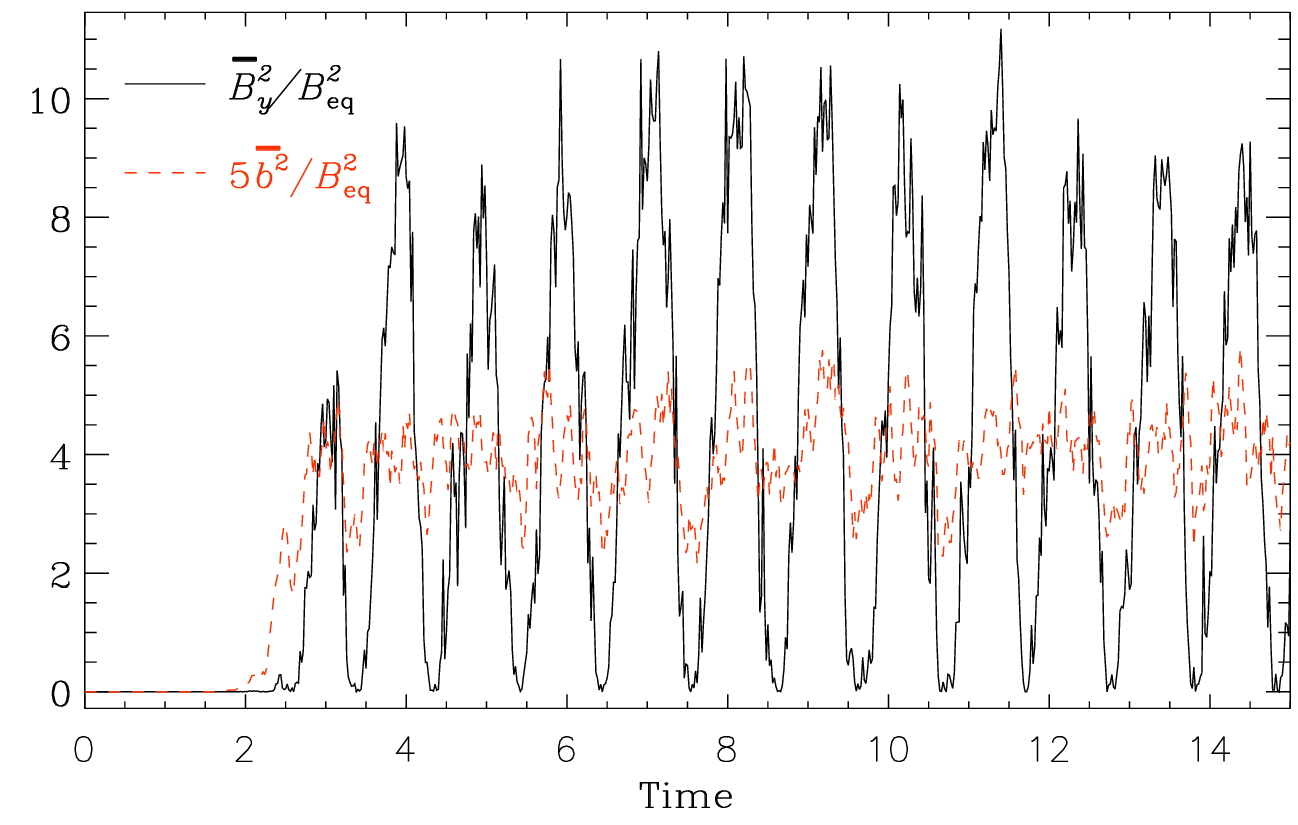} }}
    \hspace*{-0.2 cm}
     \subfloat[\centering]{{\includegraphics[scale=0.274]{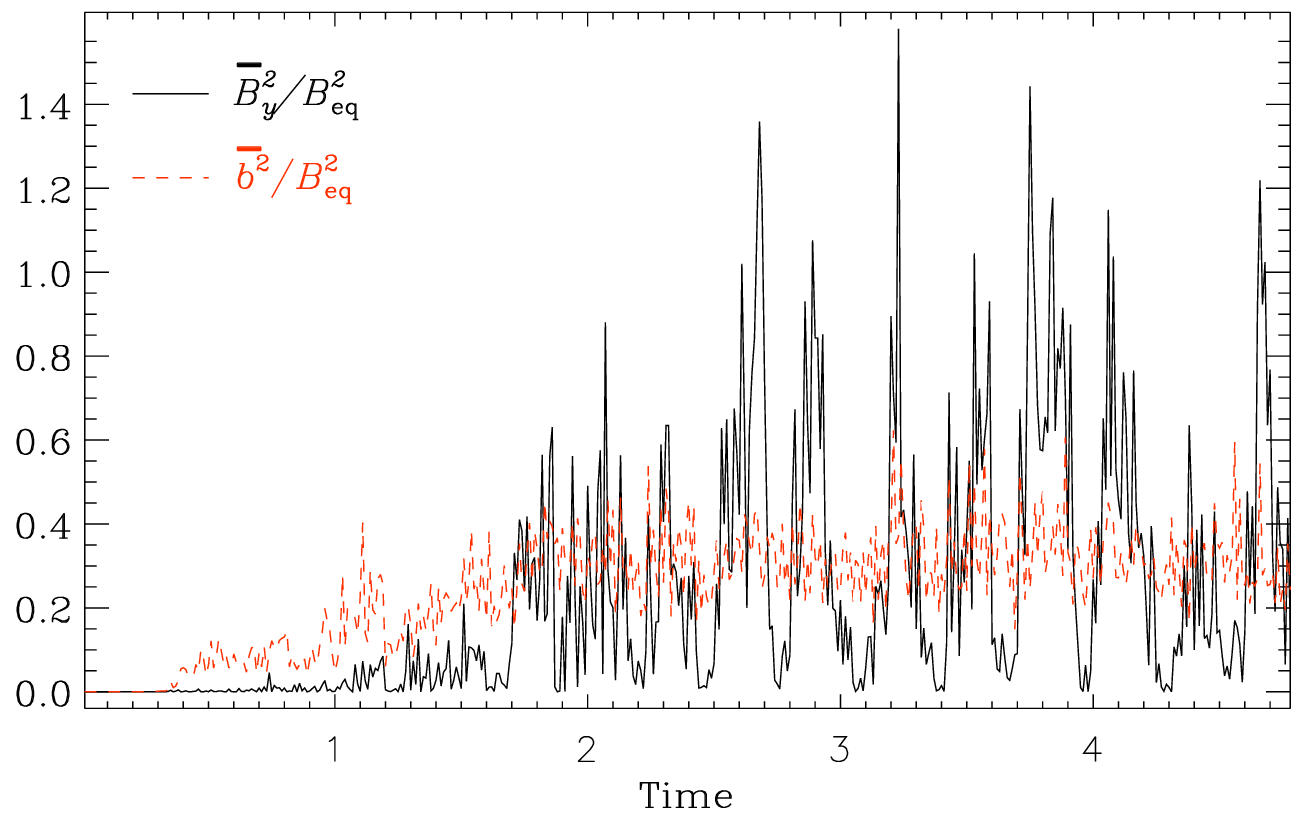} }}
    \caption{Time series of the large-scale (black/solid line) and small-scale (red/dash line) fields, when (a) large-scale dynamo is excited, and (b) both large- and small-scale dynamos are excited at the same time. 
    Results of (b) are taken from Run F at $\sigma = 0.16$ (i.e., the critical value in this setup), while in (a), the parameters are the same except the value of $k_f$ and $\nu$ are changed to 5.1 and $10^{-2}$ (Run F\textquotesingle).
    }
    \label{fig:comp}%
\end{figure}

To enhance understanding of this phenomenon, we have added a new run involving only the large-scale dynamo (similar to Run 2 of \citet{KB16}). In this run, we increased the viscosity ($\nu$) from $5 \times 10^{-3}$ to $ 10^{-2}$, to ensure that the small-scale dynamo does not operate (see Table 1). \Fig{fig:comp} provides a comparison between the scenarios where only the large-scale dynamo is evolved and where both the large-scale and small-scale dynamos are evolved together. From \Fig{fig:comp}a, it can be observed that the small-scale field is five times weaker than the large-scale field and becomes significant only when the large-scale field begins to grow. Additionally, \Fig{fig:comp}b clearly shows that the small-scale field develops significantly faster than the large-scale field. 

\begin{figure}
\centering
\includegraphics[scale=0.4]{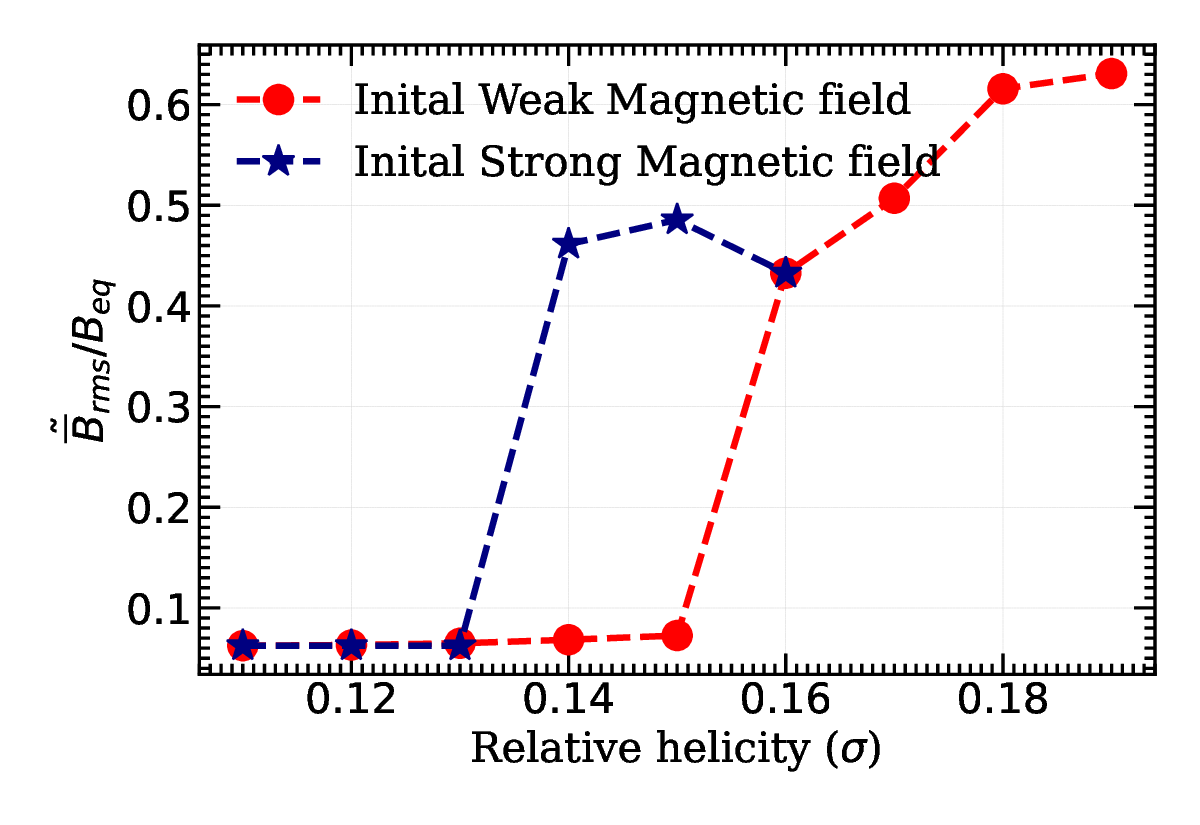} 
 \caption{Dynamo hysteresis: Variation of the temporal mean of rms value of the large-scale magnetic field normalized to $B_{eq}$ as a function of helicity parameter $\sigma$ from simulations started with a weak seed field (red filled circles) and from simulations started with a strong field of previous simulation (blue stars).}
   \label{fig:hysteresis}
\end{figure}

So far, in the previous simulations, we used a random weak seed field as an initial condition. Now, we expand our study by initiating a new simulation, where we use a snapshot from the run at $\sigma = 0.17$ (Run G) as the initial condition and introduce it into the run at $\sigma = 0.16$ (Run F). Next, we perform a sequence of simulations by systematically decreasing the value of $\sigma$, with each simulation using the output 
of the preceding one as its initial condition. Our observations unveiled the oscillatory solutions spanning a broad range of $\sigma$ values, particularly within the interval $0.13 \leq \sigma < 0.16$. Remarkably, these oscillations persisted even in scenarios where a decaying magnetic field had been present earlier. This behavior is illustrated in \Fig{fig:strong_lsd}, which showcases a snapshot of the oscillating field within the sub-critical region.

As presented in \Fig{fig:hysteresis}, a hysteresis curve is demonstrated. Within a specific range of the dynamo parameter, we identified a regime where two possible solutions coexist: a weak, decaying magnetic field and a robust, oscillatory magnetic field. The ultimate output depends heavily on the initial conditions. Notably, simulations initiated with weak magnetic fields consistently resulted in decaying solutions within this parameter range, emphasizing the system's sensitivity to its starting state. Consequently, this region exhibits bistability. It is important to emphasize that all simulations were conducted for several hundreds of diffusion times
to ensure the stability of their respective states. Therefore, our work gives additional support to the dynamo hysteresis behavior in the $\alpha\Omega$ type dynamo model, suggesting that such hysteresis is probably a characteristic feature of the $\alpha\Omega$ type solar dynamo, even in the presence of a small-scale magnetic field.

\section{Conclusions}\label{s:4}

Previous study \citep{KKB15} demonstrated the existence of two distinct dynamo modes within the subcritical region of the large-scale dynamo, i.e., there is a presence of a hysteresis behavior in the system where, depending on the initial conditions, the system exhibits bistability by accommodating both non-decaying oscillatory and decaying dynamo solutions.

Building upon this foundation, our current work takes a significant step forward by incorporating the small-scale dynamo. In real sun scenarios, both the large-scale and small-scale dynamos operate concurrently at the same location. Consequently, we have identified that the features we previously observed in the presence of only large-scale dynamo also manifest in the presence of the small-scale dynamo—thus affirming the continued existence of dynamo hysteresis.

To demonstrate this, we utilized the {\sc Pencil Code} to establish an $\alpha\Omega$ dynamo model featuring uniform shear and helically induced turbulence. Through a series of simulations conducted at varying relative helicity levels, we explored the generation of large-scale oscillatory magnetic fields in conjunction with small-scale dynamo processes.

In summary, our findings not only validate but also strengthen the conclusions drawn in our earlier work, providing a more comprehensive understanding of dynamo behavior in both large and small scales.

%
 \begin{acks}
The author wishes to extend heartfelt gratitude to Bidya Binay Karak for inspiring the initiation of this problem and for providing constant guidance and support. The author further acknowledges the anonymous referee, who provided valuable feedback and raised insightful questions, which enhances the quality of the paper. Financial support from the Department of Science and Technology (DST), India, through the Inspire fellowship, is acknowledged. The author also acknowledges the Computational support and the resources provided by the PARAM Shivay Facility under the National Supercomputing Mission, Government of India at the Indian Institute of Technology (BHU), Varanasi.
 \end{acks}

%
%
%
%
%
%
%

%
%
\bibliographystyle{spr-mp-sola}
\bibliography{paper}  
%
%
%
%

\end{document}